# Unconventional Optical Response in Engineered Au-Ag Nanostructures

Dev Kumar Thapa[1#], Subham Kumar Saha[1#], Biswajit Bhattacharyya[1#], Guru Pratheep Rajasekar[1], Rekha Mahadevu[1] and Anshu Pandey[1*]

[#] Equal Contributors

[1] Solid State and Structural Chemistry Unit, Indian Institute of Science, Bangalore 560012.

* Correspondence to anshup@iisc.ac.in

This article describes the optical properties of nanostructures composed of silver particles embedded into a gold matrix. In previous studies these materials were shown to exhibit temperature dependent transitions to a highly conductive and strongly diamagnetic state. Here we describe the anomalous optical properties of these nanostructures. Most notably, these materials fail to obey Mie theory and exhibit an unconventional resonance with a maximum at about 4 eV, while the usual gold and silver localized surface plasmon resonances are suppressed. This effect implies a significant deviation from the bulk dielectric functions of gold and silver. We further resolved this resonance into its absorbance and scattering sub-parts. It is observed that the resonance is largely comprised of scattering, with negligible losses even at ultraviolet frequencies.

**Introduction:**

Localized Surface Plasmon Resonances (LSPRs) are known to concentrate electromagnetic fields in their vicinity [1-4], potentially allowing for control and customization of electromagnetic coupling of any chromophore [5-8]. Metals such as Au and Ag are most typically employed to prepare nanocrystals for the study of visible LSPRs [9-13].

In this paper, we describe the optical properties of Au-Ag nanostructures (NS) reported in a previous work [14]. In particular, we had observed the emergence of strong diamagnetism and vanishingly small resistance in aggregates of these NS. The transition temperature to this state could be tuned by varying the mole fractions of the two metals as well as the state of ageing of the samples. It was further demonstrated that even under ambient pressures, this state could be stabilized at temperatures exceeding room temperature. These electrical and magnetic properties are anomalous for both Au and Ag and are thus an obvious outcome of nanostructuring. These anomalies motivated us to also analyse the optical properties of these materials in the context of their electrical and magnetic properties.

Figures 1a-c show the High-Annular Aperture Dark Field (HAADF) as well as elemental mapping of these NS. As shown in these figures, the NS studied in this article comprise of Ag particles embedded into an Au matrix. Unlike conventional core/shell Au-Ag structures, the size of the Ag particles used in the embedding process is small, of the order of ~1 nm, and a substantial number of Ag units are incorporated into the Au matrix of each nanoparticle. Detailed preparative

methods were described in our earlier work. As shown in the transmission electron micrograph (TEM) in Figure 1b, the samples used in this study comprise of spheroidal nanoparticles < 20 nm in diameter. Figure 1c further shows a high resolution TEM (HRTEM) image of an NS. The inset to Figure 1c shows the Fourier transform of the HRTEM image in main figure. As evident, the observed patterns are consistent with fcc lattice pattern of Au and Ag.

We now proceed to examine the optical characteristics of these NS. In ordinary metallic nanoparticles, an LSPR corresponding to the surface plasmon frequency is observed in the extinction spectrum[15]. For small, isolated particles, this corresponds to the excitation of a dipolar mode of collective electron motion. For trivial Au and Ag particles the energy of the dipolar mode is a function of the particle morphology and occurs with energy lower than ~2.3 eV and ~3.1 eV respectively for either metal [16-18]. Unlike normal Au and Ag based particles, these NS exhibit an unusual resonance with a maximum at ~4 eV (~300 nm) that occurs along with the simultaneous suppression of plasmon resonances of Au and Ag in the NS. This resonance is resolved into scattering and absorption components, and is shown to comprise almost entirely of scattering. The composition dependence of the extinction spectra of these NS is further studied. Our observations strongly deviate from expectations that arise from Mie theory and therefore point to a strong change in the dielectric constants of the materials that comprise the NS [15,19-21]. We find that this optical resonance does not appear to be the consequence of a gap in the electronic density of states.

For a nanoparticle, a resonance is obtained when the electric polarizability of the particle exhibits a pole. In the case of a spherical metal particle, this condition may be expressed as $\varepsilon' = -2n^2$[22,23]. Here $\varepsilon'$ is the real part of the nanoparticle dielectric constant and $n$ is the refractive index of the medium. Since the spectral position of a localized surface plasmon resonance (LSPR) is thus expected to vary due to the medium dielectric, these have been widely used in sensing molecular species and analytes [24-26]. While metal nanoparticles do show resonances that vary with the medium dielectric, typical shifts in spectral position are small, being only of the order of -0.2 eV per refractive index unit (RIU) [27,28]. These shifts are limited by the energy dispersion of the dielectric function $\varepsilon'(E)$ where $E$ represents the incident photon energy. In practical terms, this translates to a ~100 meV shift in the LSPR of an Au sphere when the medium is changed from air to a high index (1.5) solvent. In the case of Au and Ag nanospheres in water, resonances thus typically occur at 2.3 eV and 3.1 eV respectively.

In contrast to the well-known properties of Au and Ag nanocrystals, the NS described here exhibit several anomalies in their optical characteristics. The optical spectrum of a typical sample is shown in Figure 2. These data have been recorded by dispersing the prepared NS into deionized water. The NS were cleaned and separated from the reaction by-products prior to this step. As evident from this spectrum, the Au and Ag LSPR (labelled by arrows in Figure 2) appear to be suppressed, while a new broad feature, $R_1$ is observed. In this example, the maximum of $R_1$ appears to be at 4.15 eV (dashed arrow). We note that the precise spectral location of the maximum of $R_1$ varies somewhat from sample to sample, ranging from 3.8 eV to 4.2 eV in our studies. The peak optical extinction of $R_1$ is typically an order of magnitude greater than the peak extinction of the starting Au nanospheres (2.3 eV) used in the synthesis, thereby suggesting an extinction cross section of ~$10^{-10}$ cm$^2$. Due to its large amplitude, this feature could not be attributed to any residual molecular species in solution, and was consequently treated as a resonance associated with the NS sample. As discussed above, in the conventional Mie theoretic perspective, LSPR energies for spheres correspond to frequencies where the real part of the dielectric constant is negative [19]. In the case of both Au and Ag, the imaginary part of the dielectric constant is large at 4 eV. Further, in the ultraviolet (e.g. 4.2 eV), Ag exhibits a positive real dielectric constant (0.6), while the real dielectric constant for gold is too small (-1.47) [21]. It is thus apparent that neither of these materials is capable of supporting an LSPR at 4 eV in water (refractive index 1.36 in the ultraviolet)[29]. This analysis could either imply a strong change in the dielectric constant of the NS when compared to the dielectric constants of the constituent materials or else a non-LSPR origin to $R_1$.

We additionally note that the available structural details (Figure 1c) confirm the absence of major lattice deformations to the constituent elements, and thus this resonance is expected to have a purely electronic origin. In the context of our previously reported observations regarding vanishing resistance in these NS, it further becomes interesting to examine the lossiness of this resonance.

A quantitative decomposition of the extinction spectrum into an optical absorption and scattering spectrum was performed using an integrating sphere within an Edinburgh Instruments FLS 920 Spectrometer. Samples inside a quartz cuvette were introduced into a spectralon coated integrating sphere. These were illuminated by a variable wavelength of light (280 nm – 800 nm; 4.43 eV – 1.55 eV) generated by monochromating light from a 450 W Xenon lamp. The scattered light was collected at right angle geometry, monochromated and subsequently detected using a photomultiplier tube (Figure 3a). This enables us to directly measure the degree of light absorption into the samples and thereby resolve extinction spectra into absorption and scattering components. The optical response of a typical sample of gold nanospheres is shown in Figure 3b. In these measurements, the response of a cuvette filled with neat solvent was used to define a background response. The logarithm of the sample scatter to the solvent scatter was interpreted as the optical absorption. In the integrating sphere, light makes multiple passes through the sample, leading to enhanced absorption over the values observed in a single pass. A correction factor of 0.63 has therefore been employed to scale the optical absorption data to make these comparable to the extinction spectra. The correction factor was determined by measuring the absolute absorbance of a solution of semiconductor nanocrystals with a known optical density of 0.20 at 3.1 eV. As shown in Figure 3b, the optical extinction of Au nanospheres is largely composed of absorption. We note that these Au nanospheres serve as initial cores which are eventually converted into Au-Ag NS by deposition of additional material. Our results with respect to Au spheres are thus entirely consistent with previous reports.

The optical characteristics of Au-Ag NS are distinctly different from the properties of Au nanospheres [30]. We firstly note that exclusively elastic optical scattering was observed in the case of certain samples of Au-Ag NS. Exemplary data are plotted in Figure 3c. As apparent, the absorbance under $R_1$ is negligible, being limited only by the errors in our measurements. Further, it is evident that elastic scattering is observed over the entire visible range. Most surprisingly, even the intrinsic absorbance of Au and Ag appears to be suppressed. Figure 3d further shows that aqueous dispersions of these NS exhibit no measurable evidence of luminescence or any other form of inelastic scattering. As evident from this figure, inelastic scattering, if present would occur with a probability < 0.5% of the elastic scattering efficiency. In this particular data, the samples have been illuminated by 4.1 eV light. The feature at ~2 eV is a grating artefact.

These data clearly establish the credentials of $R_1$ as a scattering resonance. Furthermore we discount the possibility of involvement of molecular species entirely, since scattering cross sections are negligible for small molecules.

Plasmons in small (< 20 nm) metal nanocrystals are known to be lossy, and do not significantly contribute to scattering[31]. In particular, plasmons in such particles dephase rapidly (~10 fs), leading firstly to the formation of a hot electron gas that eventually dissipates its thermal energy into the lattice[32-34]. This effect becomes more pronounced in the case of smaller nanoparticles, where surface scattering is a significant contributor to plasmon dephasing [35,36]. Significant enhancement in dephasing has also been reported in the case of Au/Ag core/shell structures due to scattering of electrons at the Au/Ag interface[12]. Our observation of highly efficient light scattering from these NS is thus entirely unexpected. To further rule out measurement artefacts, we attempted to directly measure optically induced thermal transients using pump-probe spectroscopy. In these experiments, samples were illuminated by a 100 fs, 3.1 eV, 9.52 $\mu J$ pump pulses derived from a Coherent Libra amplified laser. Figure 4a shows the optical extinction spectra of the Au-Ag NS sample as well as a sample of Au nanospheres employed for this experiment. Figure 4b compares the transient response observed in the two materials. As evident from these data, no measurable response is observed in the case of Au-Ag NS (large red dots), while normal electron cooling dynamics is observed in the case of Au nanospheres. On the time resolution of our experiment, the transient response of Au nanospheres may be fit to a biexponential of 3.5ps (electron-lattice coupling) and 50 ps (lattice cooling) respectively. These data essentially confirm the absence of measurable absorbance in Au-Ag NS as well as the simultaneous strong suppression of the normal Au and Ag LSPR.

The suppression of the normal Au and Ag LSPRs as well as the emergence of a new resonance in Au-Ag NS is consistent with the reorganization of electrons within the Au-Ag NS. In the past, we have shown that appropriate nanostructuring can lead to ground state electron transfer between semiconductor nanostructures as well as the consequent emergence of new properties that are distinct from the behaviour individual materials[37,38]. The appearance of a new resonance in Au-Ag NSs thus naturally raises a question regarding its identity and origin. Towards this end, we examined the sensitivity of this resonance to its dielectric environment. For these experiments, NS were deposited onto a quartz slide. Subsequently, the extinction spectrum was measured while the quartz slide was immersed into solvents with varying refractive index. Exemplary optical spectra of the NS in air and in hexane are shown in Figure 5a. Figure 5b further shows the dependence of $R_1$ maximum on the medium refractive index. The dashed line has a -0.2 eV/RIU slope similar to the dielectric sensitivity of Au nanoparticles[28]. As evident, from these data, the spectral position of resonance maximum appears to be independent or only weakly dependent of refractive index of the medium into which the NS are immersed. Assuming that the Au-Ag NS are not completely insulated from the medium by the residual ligand shell, this could be by two different hypotheses. (1) Since $\varepsilon' = -2n^2$ no longer appears to be the resonance condition, a possible interpretation is that $R_1$ has a non-plasmonic origin, and represents a set of few or single electron transitions. Other observations, particularly its tendency to exhibit elastic light scattering appear to preclude this outcome. (2) A second possibility is that $R_1$ is a bulk-plasmon-like resonance that arises from the interior of each particle and is not a conventional LSPR. The inability of electrons to sample NS surface offers a possible explanation for the dielectric insensitivity but also suggests a significant localization of electrons within the bulk of the NS. Qualitatively, such a state bears superficial similarity to a Wigner crystal[39]. While the large estimated cross sections of $R_1$ does agree with this possibility, we note that to the best of our knowledge, such a localization effect has not been proposed or observed previously. We further note that other hypotheses may also exist that account for the observed properties.

Since $R_1$ appears peculiar to the Au-Ag NS architecture and is not formally known in the bulk form of either metal, it becomes interesting to study its response to metal overgrowth. These data are shown in Figure 6a, that shows the effects of Au overgrowth. The sample spectrum is labelled 1. Spectra have been offset for clarity. Each subsequent spectrum has been taken after the addition of 65 $\mu L$ of 1 mM $HAuCl_4$ solution in the presence of 0.1 M, 1 ml $NaBH_4$ to the sample. A gap of 10 minutes was maintained between additions. As evident from this figure, the growth of subsequent layers of Au over the Au-Ag NS leads to a gradual decrease in the amplitude of $R_1$ as well as the subsequent emergence of the usual Au plasmon resonance. It is interesting to note that negligible spectral shifts are observed in $R_1$ as well as the Au LSPR during the course of growth. The absence of a spectral shift in $R_1$ maximum upon Au coating is consistent with its insensitivity to the dielectric constant of the medium that was described in the earlier experiment. We note that such shell induced spectral shifts are well known in core LSPRs of normal metallic nanoparticles[40]. While there is indeed a spectral shift in the $R_1$ maximum immediately prior to its complete suppression (Figure 6a, spectrum 5) this shift appears well after significant Au overgrowth, and is thus distinct from a refractive index induced effect.

The properties of $R_1$ were further correlated to the mole fractions of gold ($x_{Au}$). Figure 6b, 6c and 6d show the optical spectra of three different samples of Au-Ag NS with varying Au and Ag mole fractions. For this experiment, all three samples were prepared from the same starting stock of core material. Varying amounts of Au were subsequently overgrown. Growth procedures adopted here have been described previously [14]. This led to materials with $x_{Au} = 0.80, 0.83$ and $0.87$ (Figures 6a, 6b and 6c respectively). Consistent with Figure 6a, Au growth enhances the Au LSPR while suppressing $R_1$. We attempted to better quantify the precise changes occurring in the spectra as a function of Au growth. In this regard, it is observed that the Au resonance itself (as identified through 2$^{nd}$ derivatives) does not shift measurably with Au growth. While some changes in $R_1$ position are evident, the resonance itself is too broad to enable quantification. We analysed the relative amplitude of $R_1$ (3.71 eV) and Au LSPR (2.33 eV) in these samples. It is observed that the amplitudes of the two features fluctuate rather randomly over this composition range. The lack of a trend is not unexpected, since as a scattering resonance, $R_1$ may not obey the sum rule for oscillator strengths. We were however able to find a significant relation between the width of $R_1$ and the amount of Au overgrown on the sample. We estimated the half-width of the $R_1$ resonance as the photon energy interval over which the amplitude of the $R_1$ resonance falls to half the value it has at 3.71 eV. From Figure 6f it is evident that a systematic decline in the $R_1$ width is indeed observed as $x_{Au}$ increases. As shown in our previous report, these specific samples also exhibit sharp drops in resistance when cooled below a critical temperature($T_c$) [14].

To conclude, we studied the anomalous optical characteristics of engineered Au-Ag NS. These materials exhibit a suppression of normal Mie like resonances of the constituent metals and instead, exhibit an unexpected resonance $R_1$ that exhibits a maximum in the 4 eV spectral region. The $R_1$ resonance was resolved into scattering and absorption components. It was found that in certain samples, $R_1$ has nearly perfect scattering and negligible absorption. This observation was also confirmed through pump probe experiments where electron/lattice heating from an optical pump could not be detected. It is further observed that $R_1$ does not exhibit a measurable sensitivity to the dielectric environment unlike a normal LSPR. The growth of additional Au over the NS samples leads to eventual suppression of the $R_1$ resonance and the emergence of Mie like LSPR. It is observed that $R_1$ persists at room temperature even in samples that exhibit finite electrical resistance in the ambient.

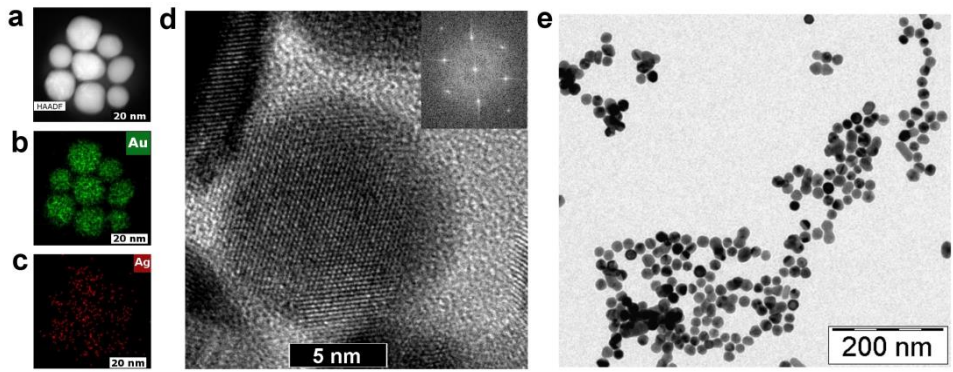

**Figure 1 a.** HAADF image of Au-Ag NS. **b.** Au and **c.** Ag distributions. **d.** HRTEM image of Au-Ag NS. Inset: Fourier transform. **e.** TEM image of ensemble of Au-Ag NS.

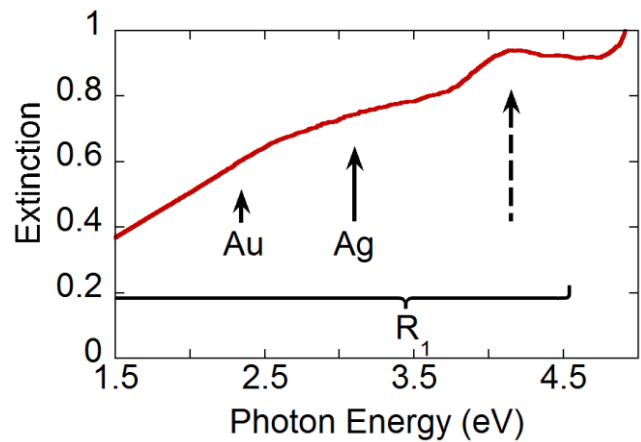

**Figure 2.** Optical extinction spectrum of Au-Ag NS. The solid arrows show the typical spectral positions of Au and Ag LSPR. The bracket indicates the range of $R_1$ as observed in this spectrum. The dashed arrow indicates the maximum of $R_1$.

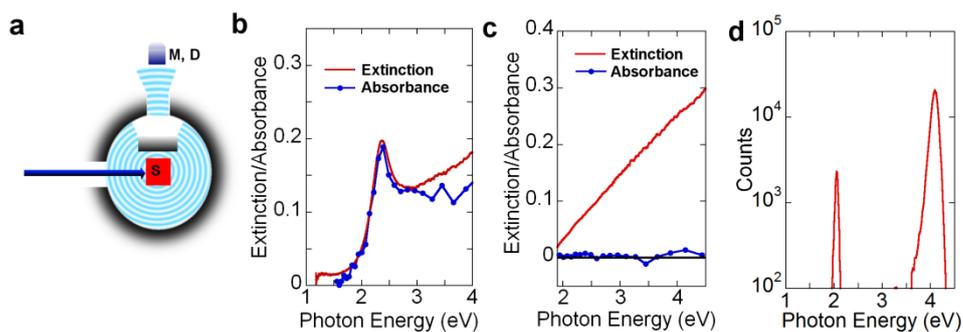

**Figure 3 a.** Schematic of absorbance measurement. Here S is the sample, M is the Monochromator and D is the detector. **b.** Extinction (red) and absorbance (blue dots) of a sample of Au nanospheres. **c.** Extinction (red) and absorbance (blue dots) of a sample of Au-Ag NS. **d.** Luminescence spectrum of the sample under 4.1 eV excitation. Only the excitation and a grating artefact are observed.

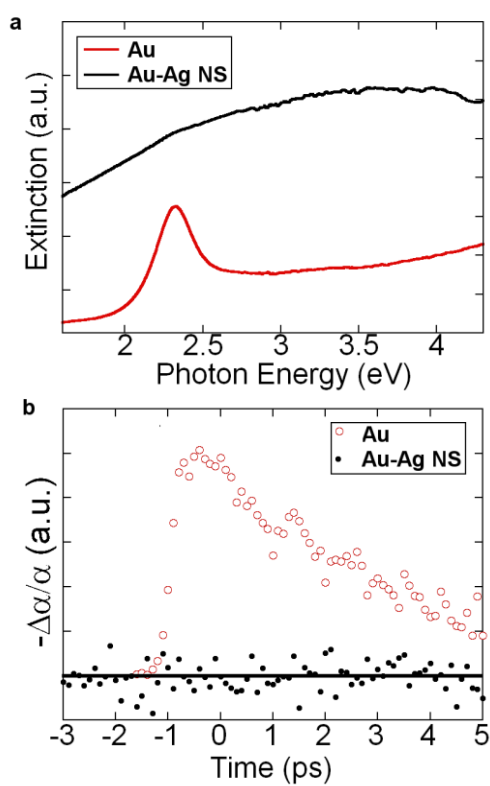

**Figure 4 a.** Optical Extinction of a sample of Au-Ag NS (black) and Au nanospheres (red). Spectra are offset for clarity. **b.** Transient bleach of the Au-Ag NS (black dots) and Au nanospheres (red circles). The horizontal black line indicates the zero of the y axis.

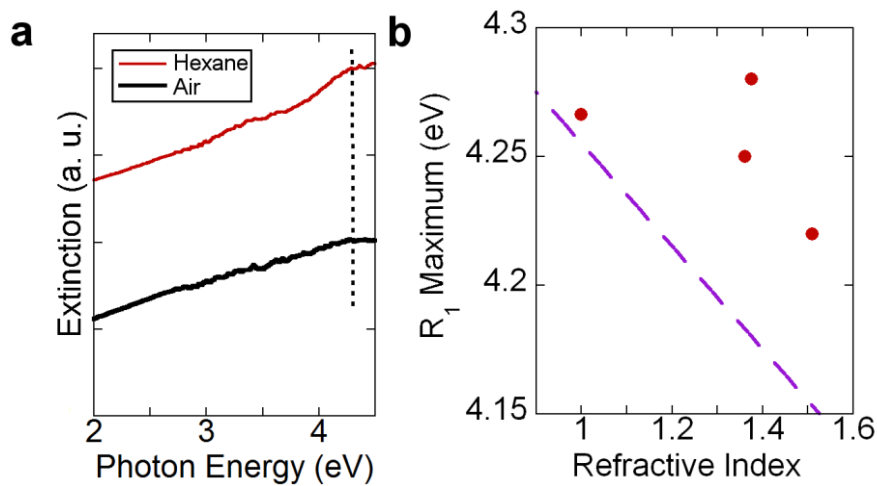

**Figure 5 a.** Optical Extinction of a sample of Au-Ag NS on a quartz slide in hexane (red) and air (black). The dashed line is a guide to the eye. **b.** $R_1$ as a function of refractive index of the medium. The dashed line corresponds to a slope of -0.2 eV/RIU that is similar to Au Nanospheres in Ref[28].

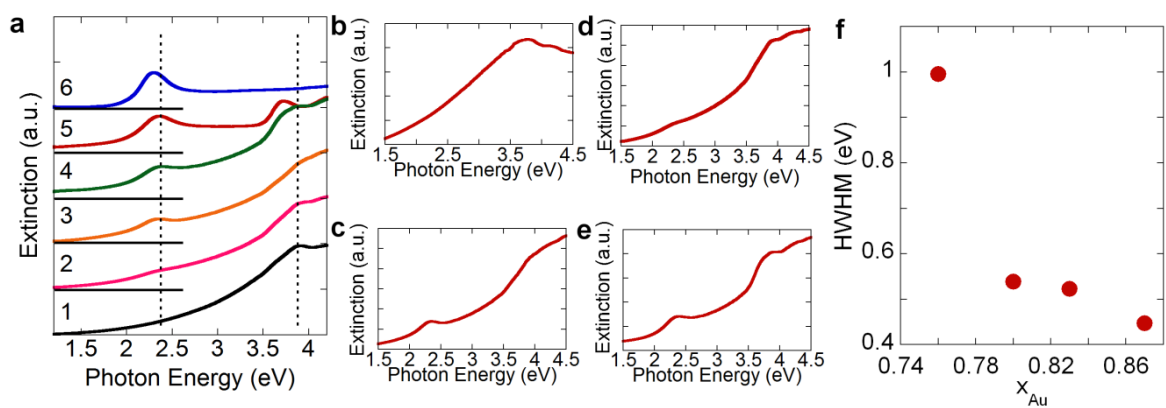

**Figure 6 a.** Optical Extinction of Au-Ag NS as a function of successive Au growth. The dashed lines are guides to the eye. **b., c., d.,** and **e.** Extinction spectra of Au-Ag NS as a function of Au shell growth in the NS. **f.** HWHM of $R_1$ as a function of $x_{Au}$.

**Acknowledgements:**

AP acknowledges financial support from the Indian Institute of Science. AP further acknowledges the use of facilities created under the DST Nanomission grant **(SR/NM/NS-1117/2012)** and DST SERB IRHPA grant (IR/S2/PU-0005/2012). We further thank CENSE for access to their facilities.

**Author contribution:**

AP conceived the idea and designed the project. AP, DKT developed initial synthetic protocols, and carried out early measurements with help from SKS. SKS, BB, and DKT devised further protocols, prepared samples and performed measurements with help from RM and GPR. AP wrote the paper with inputs from all authors.